\documentstyle[aps,preprint,floats]{revtex}
\newcommand{\Wlnu}{\mbox{$W\rightarrow l \nu$}}
\newcommand{\Wenu}{\mbox{$W\rightarrow e \nu$}}
\newcommand{\Wmunu}{\mbox{$W\rightarrow \mu \nu$}}
\newcommand{\Wtaunu}{\mbox{$W\rightarrow \tau \nu$}}
\newcommand{\Zee}{\mbox{$Z\rightarrow e e $}}
\newcommand{\Zmumu}{\mbox{$Z\rightarrow \mu \mu $}}
\newcommand{\Ztautau}{\mbox{$Z\rightarrow \tau \tau $}}
\newcommand{\Zll}{\mbox{$Z\rightarrow l l $}}
\newcommand{\MET}{\mbox{$\not\!\!E_T$}}

\newcommand{\Et}{\mbox{$E_T$}}

\def\D0{D\O}                            %D0
\def\ipb{pb$^{-1}$}                     %inverse picobarns
\def\pt{$p_T$}                          %pT
\def\met{\mbox{${\hbox{$E$\kern-0.6em\lower-.1ex\hbox{/}}}_T$ }} %missing ET
\def\d0draft{}
\def\err#1#2#3 {{\it Erratum} {\bf#1},{\ #2} (19#3)}
\def\ib#1#2#3 {{\it ibid.} {\bf#1},{\ #2} (19#3)}
\def\nc#1#2#3 {Nuovo Cim. {\bf#1} ,#2(19#3)}
\def\nim#1#2#3 {Nucl. Instr. Meth. {\bf#1},{\ #2} (19#3)}
\def\np#1#2#3 {Nucl. Phys. {\bf#1},{\ #2} (19#3)}
\def\pl#1#2#3 {Phys. Lett. {\bf#1},{\ #2} (19#3)}
\def\prev#1#2#3 {Phys. Rev. {\bf#1},{\ #2} (19#3)}
\def\prl#1#2#3 {Phys. Rev. Lett. {\bf#1},{\ #2} (19#3)}
\def\rmp#1#2#3 {Rev. Mod. Phys. {\bf#1},{\ #2} (19#3)}
\def\zp#1#2#3 {Zeit. Phys. {\bf#1},{\ #2} (19#3)}

\begin{document}
\preprint{LBL--37507}
\title{Measurements of the $W$ and $Z$ Inclusive Cross Sections and
Determination
of the \\
W Decay Width\footnote{
To be published in the Proceedings, {\em 10th} Topical Workshop on $p\bar{p}$
Collider Physics,  Fermilab, May 9--13, 1995.}$^,$\footnote{
This work was supported by
the Director, Office of Energy Research,
Office of High Energy and Nuclear Physics,
Division of High Energy Physics of the U.S.~Department of Energy
under Contract DE-AC03-76SF00098.}}

\author{Anthony L. Spadafora\footnote{Present address: Center for
Particle Astrophysics, University of California, Berkeley,
CA 94720}}
\address{Lawrence Berkeley Laboratory \\}
%%%.............. DRAFT \today .............}

\maketitle

\begin{abstract}
Recent results on the production of $W$ and $Z$ gauge bosons
in $p\bar{p}$ collisions at $\sqrt{s}=1.8$~TeV
at the Fermilab Tevatron Collider from the \D0 and CDF experiments
are reviewed.
Measurements of the inclusive cross sections times leptonic branching
ratios in both the electron and muon decay channels are summarized and compared
to QCD predictions.
Using the ratio
$R = {\sigma_W \cdot B(W\rightarrow l\nu)}/
{\sigma_Z \cdot B(Z\rightarrow ll)}$ and
assuming standard model couplings, an indirect determination of the
W decay width is obtained.
By comparing this measured value with the predicted value for the $W$ width,
a limit on the deviation from the standard model is obtained.
\end{abstract}

%----------------------------------------------------------------------
The production cross sections times leptonic branching ratios
of $W$ and $Z$ gauge bosons are one of the fundamental measurements that can
be performed at a hadron collider.
These measurements are of interest in their own right
and, in addition, the ratio
$R \equiv {\sigma_W \cdot B(W\rightarrow l\nu)}/
{\sigma_Z \cdot B(Z\rightarrow ll)}$ provides a
measurement of the $W$ decay width.
While the $Z$ boson properties (including its decay width) have been
measured to great precision at LEP,  hadron colliders provide
the only means of measuring the basic properties of the $W$ boson at present.

In this paper, I summarize the recent results obtained by
the \D0\ and CDF experiments
on the production of $W$ and $Z$ bosons.
Decays to final states including electrons and muons
have been analyzed.
The data samples used here are from the 1992--3 run of the Tevatron collider
(Run 1a) for which the integrated luminosity was $\sim 13$~\ipb\ and
$~\sim 20$~\ipb\ for \D0 and CDF, respectively.
Preliminary results are also presented from  \D0  using a
partial data sample from the 1994--95 run.

%----------------------------------------------------------------------
\section*{The Cross Section Measurements}
At $\sqrt{s}=1.8$~TeV, vector boson production in  $p\bar{p}$ collisions
proceeds primarily via $q\bar{q}$ annihilation accompanied by the
emission of gluons.
In the cross section measurements described here there are no
requirements on the
jets or transverse momentum of the vector boson, so we are integrating
over all orders of QCD processes.
At this center of mass energy, processes such
as $t\bar{t}$ or vector boson pair production contribute about a factor
of a thousand less than $q\bar{q}$ annihilation  to the inclusive cross
section.
Absolute predictions for $\sigma_W$ and $\sigma_Z$ have been calculated
to order $\alpha_s^2$ by van Neerven {\em et al.}\cite{HVM}.
The first order correction to the Born term is approximately
a 20\% increase; the change at the second order is approximately a
2\% increase.
The ratio $\sigma_W/\sigma_Z$, however, changes by only $\sim 0.6\%$ in
going from the Born approximation to the second order calculation.
The major source of uncertainty in the calculation is due to the
choice of parton distribution function (pdf).

\subsection*{Event Selection and Data Analysis}
The $W$  boson inclusive cross section is calculated as
\begin{equation}
\sigma_W \cdot B({\mbox{$ W\rightarrow l \nu$}}) =
\frac{N_{obs}-N_{bkgd}}{ {\cal A}_W \cdot \epsilon_W \cdot {\cal L}}
\label{eq_cs}
\end{equation}
where $N_{obs}$ is the observed number of candidate events,
$N_{bkgd}$ is the calculated number of expected background events,
${\cal A}_W$ is the kinematic and geometric acceptance,
$\epsilon_W$  is the detection efficiency, and
$\cal{L}$ is the integrated luminosity used in the analysis.
The $Z$ boson cross section, $\sigma_Z\cdot B(\Zll)$,
is calculated in a similar fashion.
In computing the cross section ratio, the luminosity
and part of the systematic error cancels:
\begin{equation}
R = {{\sigma_W \cdot B(W\rightarrow l\nu)}\over
{\sigma_Z \cdot B(Z\rightarrow ll)}} =
\frac{N_W {\cal A}_Z \epsilon_Z}{N_Z {\cal A}_W \epsilon_W}
\label{eq_R}
\end{equation}
where $N_W(N_Z)$ is the background corrected number of $W(Z)$ candidates.

In \D0\cite{D0_NIM}, electrons are detected in hermetic,
uranium liquid-argon calorimeters with an energy resolution of
about $15\%/\sqrt{E (\mbox{GeV})}$.
The central and end calorimeter regions are used in both the $W$ and $Z$
analyses, covering pseudorapidity ($\eta$) range:
$|\eta|<1.1$ and $1.5 < |\eta| < 2.5$ respectively.

Muons are detected as tracks in three layers of proportional
drift tube chambers outside the calorimeter: one 4-plane layer
is located inside a magnetized iron toroid and two 3-plane layers are
located outside.
The muon momentum resolution is
$\sigma(1/p) = 0.18(p-2)/p^2~\oplus~0.008$ (with $p$ in GeV/c).
Muons that passed through the central
iron toroid ($|\eta|<1.0$) were used in the analyses described here.

Neutrinos are inferred from the observed missing transverse energy
(\MET) which is calculated using all the energy detected in the calorimeter
cells out to pseudorapidity of $4.2$.
For electron channel decays, the \MET\ resolution is dominated
by the underlying event and is $\sim 3$~GeV.
%In a study of minimum bias events, the resolution on \MET\ is
%found to be $\sigma(\MET) = 1.08\; \rm GeV + 0.019\times(\Sigma\Et)$.
For the muon channel decays, the muon transverse momentum is added to
the calorimeter energy to calculate the total \MET,
and the muon momentum resolution dominates the \MET\ resolution.

The \D0 $W$ and $Z$ electron channel analyses\cite{D0_WZ_XSEC}
 base their event selection
on a sample obtained with a single electron trigger ($Et > 20$~GeV).
Offline, it is required that there be at least one electron with $\Et > 25$~GeV
that passes ``tight'' electron identification cuts.
Details of the electron identification are given in Ref.~\cite{D0_TOP_PRD},
with the main features being an electromagnetic (EM) cluster in the calorimeter
with a matching track in the central tracking chambers.
The electron is required to be isolated, with isolation fraction $I < 0.1$.
The isolation variable is defined as
$\it{I}$=($E_{\rm tot}$(0.4)-$E_{\rm EM}$(0.2))/$E_{\rm EM}$(0.2),
where $E_{\rm tot}$(0.4)
is the total calorimeter energy inside a cone of radius $\sqrt{\Delta\eta^2
+\Delta\phi^2} = 0.4$ and $E_{\rm EM}$(0.2) is the electromagnetic energy
inside a cone of 0.2.
The cluster is also required to have transverse and longitudinal shapes
consistent with
those expected for an electron based on test beam measurements and
Monte Carlo simulations.

To select \Wenu\  candidates,  in addition
to the ``tight" electron with $\Et > 25$~GeV,
events are required to have  missing transverse energy $\MET > 25$~GeV.
To select \Zee\  candidates,  in addition
to the ``tight" electron with $\Et > 25$~GeV,
events are required to have a second electron with $\Et > 25$~GeV
but the electron identification requirements are loosened
by not requiring the track match in order to increase the efficiency.
The invariant mass of the electron pair is required to be in the
range $75 < M_{ee} < 105$~GeV/$c^2$.

In an analysis of the 1992--93 data sample, corresponding to
$12.8 \pm 0.7$~pb$^{-1}$, 10338 $W$ and 775 $Z$ candidate events were found.
The mass spectra for the \Wenu\ and \Zee\ events are shown in
%Fig.~\ref{fig_d0run1a}.
Fig.~1.

The \D0 muon channel $W$ and $Z$ analyses use an event sample
that fired a  single muon trigger that had a threshold of
$p_T > 15$~GeV.
Offline, the events were required to have a reconstructed muon with
$p_T > 20 $~GeV.
For \Wmunu\ events, the missing transverse energy was required to be
$\MET > 20$~GeV.
For \Zmumu\ events, the offline threshold on the second muon was lowered
to 15~GeV and the muon identification criteria were loosened.

The main features of the D\O\ muon identification
(see Refs.~\cite{D0_WZ_XSEC,D0_TOP_PRD} for details)
include a good quality muon track that has a calorimeter confirmation
signal and has a stringent match with a track
in the central detector.
%%%%%%%%%%%%%%%%%%%%%%%%%%%%%%%%%%%%%%%%%%%%%%%%%%%%%%%%%%%%%%%%%%%%%%%
% has 1) a calorimeter confirmation ($\Et > 1$~GeV),
% 2) a path that traverse a minimum field integral of 2.0 T$\cdot$m,
% 3) a stringent track match with a track in the central detector,
% 4) a good quality global fit with the vertex and a central detector track,
% 5) a muon time of origin within 100 ns of the beam crossing,
% 6) energy in the calorimeter consistent with single muon ionization
% within a cone of radius $\sqrt{\Delta\eta^2+\Delta\phi^2}=0.2$
% and with less than 6 GeV of additional energy in a cone of 0.6, and
% 7) a good muon impact parameter.
%%%%%%%%%%%%%%%%%%%%%%%%%%%%%%%%%%%%%%%%%%%%%%%%%%%%%%%%%%%%%%%%%%%%%%%%
Cosmic ray background was reduced
by rejecting muons that also had hits or tracks
within $10^\circ$ in $\theta$ and $20^\circ$ in $\phi$
in the muon chambers on the opposite side of the interaction point.
For the \Wmunu\ selection, events that were \Zmumu\  candidates were removed.

{}From an analysis of the 1992--93 data sample, corresponding to
$11.4 \pm 0.6$~pb$^{-1}$, 1665~$W$
and 77~$Z$ candidate events were found.
The observed mass spectra for the \Wmunu\ and \Zmumu\ events are shown
in
%Fig.~\ref{fig_d0run1a}.
Fig.~1.

A preliminary analysis of a partial sample of the 1994--95 data, using the same
requirements as
described here, corresponding to $25.1\pm 1.4$~\ipb, yielded 20998 \Wenu\
and 1634 \Zee\ candidates;
an analysis of $30.7\pm 1.7$\ipb\ yielded 4516 \Wmunu\
 and 168 \Zmumu\ candidates.
The spectra are shown in
%Fig.~\ref{fig_d0run1b}.
Fig.~2.

In CDF\cite{CDF_R_PRL,CDF_R_PRD}, central electrons
are detected in a lead--scintillator EM calorimeter that covers
the rapidity range $|\eta| < 1.05$
with energy resolution of $13.5\%/\sqrt{E (\mbox{GeV})}$.
Forward electrons are detected in lead--proportional tube calorimeters
that cover
$1.1 < |\eta| < 2.4 $  (plug)
and  $2.4 < |\eta| < 4.2 $  (forward),
with an energy resolution of $28\%/\sqrt{E (\mbox{GeV})}$ (plug)
and $25\%/\sqrt{E (\mbox{GeV})}$ (forward).
Neutrinos are identified using the missing transverse energy (\MET) in the
event, where \MET\ is the magnitude of the vector
sum of all calorimeter tower transverse energies using towers $|\eta| < 3.6$.
For $W$ events, where the neutrino \pt\ is of order 20 -- 40 GeV,
the resolution on \MET\ is $\sim 3$~GeV.

The electron channel $W$ and $Z$ analyses are based on a common set of
inclusive electrons with $\Et > 20$~GeV in the central region.
It is required that the event was triggered by a central electron.
Tight selection criteria\cite{CDF_R_PRD} are placed on this first,
central electron
including an isolation cut of $ I < 0.1$ where $ I$ is defined as
$ I = (\Et^{Cone} - \Et^{Cluster})/ \Et^{Cluster}$, and
$\Et^{Cone}$ is the total transverse energy in a cone of 0.4 and
$ \Et^{Cluster}$ is the EM transverse energy in the electron cluster.

$W$ candidates are selected from the inclusive electron sample by requiring
that the event is not a $Z$ candidate and that the missing transverse energy of
the event be $\MET > 20$~GeV.
To select $Z$ candidates, a second electron with looser identification criteria
is required and is not restricted to the central region but can be in the plug
or forward regions. The energy of this second electron is required to be
	$\Et > 20$~GeV if in the central,
$\Et > 15$~GeV if in the plug, or
$\Et > 10$~GeV if in the forward region.
The invariant mass of the electron pair was required to be in the range
$66  < M_{ee} < 116$~GeV/c$^2$.

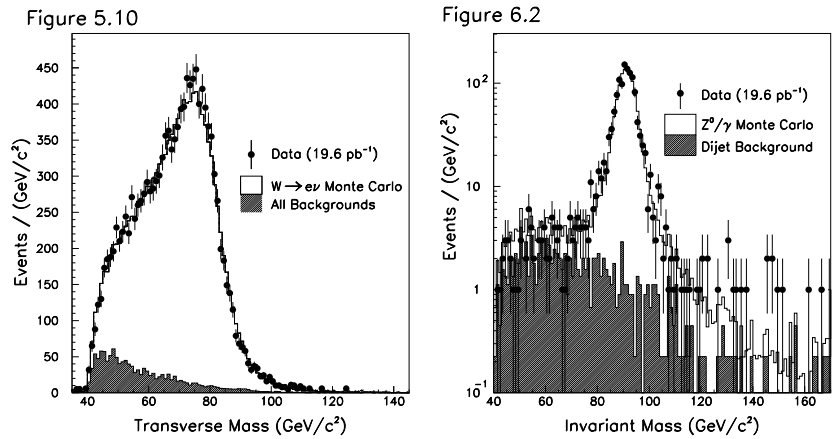
The CDF 1992--93 data sample, which corresponds to $19.6 \pm  0.7$~\ipb,
yielded 13796 \Wenu\ candidates and 1312 \Zee\  candidates; the
spectra are shown in
%Fig.~\ref{fig_cdfspectra}.
Fig.~3.

The total backgrounds estimated for these event samples are shown in the
spectra as hashed areas in
%Fig.~\ref{fig_d0run1a}
Fig.~1
and
%Fig.~\ref{fig_cdfspectra}
Fig.~3.
and are listed as a percentage of the observed number of events in
Table~\ref{tab_bkgd}.
%----------------------------- BEGIN TABLE ------------------------------
\begin{table}
\caption{Estimates of Backgrounds}
\label{tab_bkgd}
\begin{tabular}{lcccc}
D0 1992--93 data   &   \Wenu   & \Zee  & \Wmunu   & \Zmumu \\
\hline
$N_{\rm obs}$ & $ 10338 $ & $   775 $ & $  1665 $ & $ 77 $ \\ \cline{2-5}
Backgrounds(\%): & & & & \\ \hspace{.5 cm} Multijet         & $3.3 \pm 0.5 $
& $2.8 \pm 1.4 $    & $5.1 \pm 0.8      $ & $2.6 \pm 0.8 $ \\ \hspace{.5 cm}
$Z\rightarrow ee, \mu\mu, \tau\tau$   & $0.6 \pm 0.1 $    & ---   & $7.3 \pm
0.5 $                & $0.7 \pm 0.2 $ \\ \hspace{.5 cm} \Wtaunu          & $1.8
\pm 0.1 $    & ---   & $5.9 \pm 0.5 $                & --- \\ \hspace{.5 cm}
Cosmic/Random    &  ---              & ---   & $3.8 \pm 1.6 $                &
$5.1 \pm 3.6 $ \\ \hspace{.5 cm} Drell-Yan        &  ---              & $1.2
\pm 0.1 $            & ---       & $1.7 \pm 0.3 $ \\ \cline{2-5} Total
Background(\%)                 & $ 5.7 \pm 0.5 $   & $ 4.0 \pm 1.4 $ & $22.1
\pm 1.9 $ & $10.1 \pm 3.7 $ \\ \hline CDF 1992--93 data & {\Wenu}
&   {\Zee}  & &               \\ \hline  $N_{\rm obs}$   &13796
&    1312           &&\\ \cline{2-3} Backgrounds(\%) & & & & \\  \hspace{.5cm}
Multijet         & $ 6.5 \pm 1.1 $    &  $ 1.5\pm 0.7$    &&\\ \hspace{.5cm}
\Zee             & $ 2.0 \pm 0.3 $    &  $ - $            &&\\ \hspace{.5cm}
\Wtaunu          & $ 3.4 \pm 0.2 $    &  $ - $            &&\\ \hspace{.5cm}
\Ztautau         & $ 0.4 \pm 0.1 $    &  $ 0.1 \pm 0.1 $  &&\\ \hspace{.5cm}
Drell-Yan        &  ---               &  $0.5  \pm 0.2 $  &&\\ \cline{2-3}
Total Background(\%)               & $ 12.3 \pm 1.2 $   &  $2.1  \pm 0.7 $   &
&  \end{tabular} \end{table}
%----------------------------- END TABLE ---------------------------------

A major background to all the analyses is from QCD
multijet events where a jet fluctuated so as to pass the lepton criteria.
The details of how this is estimated in each analysis vary but the
basic idea is to estimate the  number of events in a background--dominated
sample, such as non--isolated electrons,
and extrapolate this into the signal region.
The contamination in both the $W$ and $Z$ samples arising from
\Wtaunu\ and $Z\rightarrow \tau \tau$ decays is estimated
by Monte Carlo.
The $W$ samples also contain  background from \Zll\
decays where one lepton is missed or misidentified.
The \D0 muon channel analyses have a residual cosmic ray background
(Table~\ref{tab_bkgd}).
Finally, in determining the \Zll\ cross section,  a correction
(which is listed as a background in Table~\ref{tab_bkgd}) is made for the
Drell--Yan process where the lepton pair is produced via a virtual photon.
This correction is sensitive to the choice of $Z$ mass window.

The combined  kinematic and geometric acceptance for these analyses
(Table~\ref{tab_analyses})
are calculated by Monte Carlo, using a parton level generator.
%----------------------------- BEGIN TABLE ------------------------------
\begin{table}
\caption{Analysis results}
\label{tab_analyses}
\begin{tabular}{lcccc}
\D0 1992--93          &   \Wenu                 &   \Zee              & \Wmunu
& \Zmumu \\                 \hline Nobs    &   10388                         &
775               &   1665            & 77 \\                    Background
(\%)     &$    5.7 \pm 0.4     $&$  4.0 \pm 1.4    $ &$  22.1 \pm 1.9    $&$
10.1 \pm 3.7     $\\  Acceptance (\%)     &$    46.0 \pm 0.6    $&$  36.3 \pm
0.4   $ &$  24.8 \pm 0.7    $&$  6.5 \pm 0.4     $\\   Efficiency (\%)     &$
70.4 \pm 1.7    $&$  73.6 \pm 2.4   $ &$  21.9 \pm 2.6    $&$  52.7 \pm 4.9
$\\  Integrated L (\ipb) &$ 12.8 \pm 0.7       $&$ 12.8 \pm 0.7  $   &$ 11.4
\pm 0.6     $&$ 11.4 \pm 0.6  $ \\         \hline \D0 1994--95 {\em
(Preliminary)}         &   \Wenu                 &   \Zee              & \Wmunu
& \Zmumu \\                 \hline Nobs    &   20988                     &
1634           & 4516            &    168\\                    Background (\%)
&$ 17.3 \pm 2.2  $&$ 11.0 \pm 2.4 $ &$ 17.3 \pm 1.1    $&$  10.1 \pm 3.7 $\\
Acceptance (\%)     &$ 46.1 \pm 0.6  $&$ 36.3 \pm 0.4 $ &$ 22.0 \pm 0.9    $&$
5.1 \pm 0.6  $\\   Efficiency (\%)     &$ 66.9 \pm 4.1  $&$ 70.6 \pm 4.6 $ &$
28.6 \pm 1.9    $&$  60.9 \pm 2.6 $\\  Integrated L (\ipb) &$ 25.1 \pm 1.4  $&$
25.1 \pm 1.4 $ &$ 30.7 \pm 1.7    $&$  30.7 \pm 1.7 $\\ \hline CDF 1992--93
& \Wenu                   &\Zee  & &  \\ \cline{1-3} Nobs	            &
13796    	        &   1312 &&\\ Background (\%)     &   $12.3  \pm 1.2 $
&    $   2.1  \pm 0.7 $  &&\\ Acceptance (\%)	    &   $34.2 \pm 0.8 $    &
$   40.9 \pm 0.5 $  &&\\ Efficiency (\%)	    &   $72.0 \pm 1.3 $    &
$   69.6 \pm 1.7 $  &&\\ Integrated L (\ipb) &   $19.6  \pm 0.7$    &    $
19.6 \pm 0.7 $  && \end{tabular} \end{table}
%----------------------------- END TABLE ---------------------------------
The transverse momentum distribution of the $W$ and $Z$ bosons are simulated
in the \D0\ analyses using the NLO calculation of Arnold and
Kauffman\cite{AK},
while in the CDF analysis  the MC 4-vectors are given
the \pt distribution observed in the data.
The 4-vectors are then run through a fast detector simulation which
models the detector fiducial volume as well as the various resolutions.
The largest contribution to the systematic error
in the acceptance (Table~\ref{tab_analyses}) arises
from the choice of pdf. Other errors included are from varying the $W$ mass,
the simulation of the \pt($W$) and \pt($Z$) distributions, radiative
corrections,  the detector simulation of the missing \Et\ distributions and the
detector energy  scale.
In computing the ratio of the acceptances, ${\cal A}_W/{\cal A}_Z$, part of
the systematic  errors cancel. The ratios obtained are:
$1.26 \pm 0.013$ (\D0 electron),
$3.82 \pm 0.22$  (\D0 muon), and
$0.835 \pm 0.013$ (CDF electron).

The net detection efficiency (Table~\ref{tab_analyses})
includes both the trigger and offline
efficiencies. These are estimated from the data using \Zll\ events
since the trigger required only one lepton.
In \D0, the electron channel trigger is found to be
$\sim 95\%$ efficient; the muon trigger efficiency
is $40$\% ($70$\%) efficient for $W$($Z$) boson events.
The CDF efficiencies in Table~\ref{tab_analyses} include a factor of $(0.955
\pm 0.011)$  which is the efficiency of the requirement that the primary vertex
of the
event be within 60~cm of the nominal interaction point.
The systematic error partially cancels in forming the efficiency ratio,
$\epsilon_W/\epsilon_Z$, and the values obtained are:
$0.957 \pm 0.017$ (\D0 electron),
$0.416 \pm 0.023$ (\D0 muon), and
$1.035 \pm 0.016$  (CDF electron).

In both experiments, the luminosity is measured by scintillator hodoscopes.
\D0\cite{D0_LUM} uses its Level 0 trigger hodoscope at $z = \pm 1.4$~m.
The  north-south
coincidence rate is measured and  corrected for multiple interactions.
The visible cross section is calculated to be $\sigma_{L\O} = 46.7 \pm 2.5$~mb,
which results in a 5.4\% relative error on the luminosity determination.
This calculation is based on an average of the published
CDF\cite{CDF_LUM} and E710\cite{E710_LUM} measurements of the
total, elastic, and single diffractive cross sections, with
the MBR\cite{MBR} and Dual Parton Model DTUJET-93\cite{DTUJET} Monte Carlo
routines used to determine the hodoscope acceptance.
CDF uses for its luminosity measurement its BBC scintillator planes at
$z = \pm 5.8$~m.
The visible cross section, based on the CDF measurement\cite{CDF_LUM} is
$\sigma_{BBC} = 51.15 \pm 1.6 $~mb, which yields a 3.6\% relative error on
the luminosity determination.

The resulting cross sections, which are
calculated using Eq.~\ref{eq_cs}, are listed in
Table~\ref{tab_cs},  where the first error given is statistical  and
the second is the total systematic error, including the luminosity.
%----------------------------- BEGIN TABLE ------------------------------
\begin{table}
\caption{Cross Section Results for electron $(e)$, muon $(\mu)$, and combined
$(e+\mu)$ channels. When two errors are given the first is the statistical
error
and the second is total systematic error.}
\label{tab_cs}
\begin{tabular}{llll} & $\sigma_W\cdot B(W^\pm\rightarrow l^\pm \nu)\; \rm (nb)
$ & $\sigma_Z\cdot B(Z\rightarrow l^+l^-) \;\rm (nb)   $   &
\multicolumn{1}{c}{$R$}  \\ \hline \bf{1992--93} &   &   \\ \D0 $(e)$       &$
2.36 \pm 0.02 \pm 0.15 $&$ 0.218 \pm 0.008 \pm 0.014 $&$ 10.82 \pm 0.41 \pm
0.30 $\\ \D0 $(\mu)$     &$ 2.09 \pm 0.06 \pm 0.25 $&$ 0.178 \pm 0.022 \pm
0.023 $&$ 11.8^{+1.8}_{-1.4}  \pm 1.1 $\\ \D0 $(e+\mu)$   &$     $&$
$&$ 10.90 \pm 0.49  $\\ CDF (e)         &$ 2.51 \pm 0.12          $&$ 0.230 \pm
0.012           $&$ 10.90 \pm 0.32 \pm 0.29 $\\ \hline
\multicolumn{2}{l}{\bf{1994--95} {\em (Preliminary)} }  &   \\ \D0 (e)
&$ 2.24 \pm 0.02 \pm 0.20 $&$ 0.226 \pm 0.006 \pm 0.021$&$ 9.9 \pm 0.3 \pm
0.8$\\ \D0 $(\mu)$     &$ 1.93 \pm 0.04 \pm 0.20 $&$ 0.159 \pm 0.014 \pm
0.022$&$ 12.3 \pm 1.1 \pm 1.2$\\ \hline \bf{1988--89}    &   &   \\ CDF (e)
&$ 2.19 \pm 0.04 \pm 0.21 $&$ 0.209 \pm 0.013 \pm 0.017$&$ 10.2 \pm 0.8 \pm
0.4$\\ \hline \bf{Standard}   &  &   &   \\ \bf{Model}      & $
2.42^{+0.13}_{-0.11}  $ & $0.226^{+0.011}_{-0.009}$  &    \end{tabular}
\end{table}
%----------------------------- END TABLE ---------------------------------
These values are compared to the theoretical prediction
(taken from Ref.~\cite{D0_WZ_XSEC}) in
%Fig.~\ref{fig_cs}.
Fig.~4.
The total cross sections are calculated to be
$\sigma_W = 22.35$~nb and $\sigma_Z = 6.708$~nb using a numerical
calculation program from
Ref.~\cite{HVM} and using the CTEQ2M pdf~\cite{CTEQ},
$M_Z=91.19$ GeV/c$^2$~\cite{PDG},
$M_W = 80.23 \pm 0.18$ GeV/c$^2$~\cite{WMASS}, and
$\sin^2\theta_W \equiv 1-(M_W/M_Z)^2 = 0.2259$.
The branching ratios used are  $B(\Wlnu) = (10.84 \pm 0.02)\%$
(calculated following Ref.\cite{RWT} but with the above $M_W$), and
$B(Z\rightarrow ll) = (3.367 \pm 0.006)\%$~\cite{PDG}.
The width of the band in
%Fig.~\ref{fig_cs}
Fig.~4.
indicates the error in the predicted value, due primarily
to the choice of pdf (4.5\%) and to the use of a NLO pdf with the van Neerven
{\em et al.} NLLO calculation\cite{HVM} (3\%).

\section*{The $W$ Decay width}
The ratio of these cross sections can be used to obtain an indirect
measurement of the $W$ leptonic branching ratio, B(\Wlnu),  and the W
total decay width, $\Gamma_W$. Using the measurement of R (Eq.~\ref{eq_R}),
we obtain a measurement of the $W$ leptonic branching ratio:
\begin{equation}
	B(\Wlnu) = \frac{B(\Zll) }{\sigma_W/\sigma_Z}\times R
\label{eq_BR}
\end{equation}
where the factor multiplying $R$ is computed from quantities
independent of this measurement.
$B(\Zll)$ is measured at {\sc LEP} to be $(3.367 \pm 	0.006)\%$\cite{PDG}.
The predicted ratio of total cross sections $\sigma_W/\sigma_Z$
has been calculated using the procedure and parameters
given above to be $\sigma_W/\sigma_Z = 3.33 \pm 0.03$.
The  uncertainty in this ratio is dominated by the choice of pdf;
the  1\% error given here covers the variation obtained when using various
CTEQ2 and MRS pdf's.

By further assuming the standard model partial width $\Gamma(\Wlnu)$,
we obtain a value for the $W$ total decay width:
	\begin{equation}
\Gamma_W =  \frac{\Gamma(\Wlnu)}{B(\Wlnu)} =
\frac{\Gamma(\Wlnu)\cdot \sigma_W/\sigma_Z}{B(\Zll)}\times \frac{1}{R}.
\label{eq_GW}
\end{equation}
The standard model $W$ partial width is given\cite{RWT} by:
\begin{equation}
\Gamma(\Wlnu) = \frac{G_F}{\sqrt{2}} \frac{M_W^3}{6\pi}(1+ \delta)
\end{equation}
where $\delta \sim -0.35\%$; there are no ``oblique'' corrections (i.e. through
loops of new particles) beyond those to $M_W$ itself.
Using the value for $M_W$ given above, this yields
$\Gamma(\Wlnu) = 225.2 \pm 1.5$~MeV.
(In computing $\Gamma_W$ from Eq.~\ref{eq_GW}, the correlation of the
error on $\Gamma(\Wlnu)$ with that on $\sigma_W/\sigma_Z$ through their
common dependence on  $M_W$ is taken into account in the results given below.)

The values of  the experimental ratio $R$ are listed in Table~\ref{tab_cs}.
\D0 combines its results from the electron and muon decay channels to
obtain a combined value for $R$.
Using Eq.~\ref{eq_BR}, this yields  a measurement of the leptonic
branching ratio of $B(\Wlnu) = (11.02 \pm 0.50)\%$ and, using
Eq.~\ref{eq_GW}, yields a measurement of the $W$ width of
$\Gamma_W = 2.044 \pm 0.092$~GeV.

{}From analysis of their electron channel data, CDF obtains
\cite{CDF_R_PRL,CDF_R_PRD}  a value of
R (Table~\ref{tab_cs}) which yields a branching ratio of
$B(\Wenu) = (10.94 \pm 0.33 \pm 0.31)\%$.
{}From this they obtain
a measurement of the $W$ width of
$\Gamma_W = 2.064\pm  0.061 \pm 0.059$~GeV, where the input quantities
used differed slightly from those given above.
Using the the above input quantities and the CDF measured value of $R$
yields  $\Gamma_W = 2.043 \pm 0.082$~GeV.
Assuming the values of $R$ measured by each experiment are independent,
they can be combined to give $R = 10.90 \pm 0.32$.
Using the above input quantities this yields $\Gamma_W = 2.043 \pm
0.062$~GeV, where the $3.0\%$ error predominately comes from the
experimental error on $R$.

It is of interest to compare and combine this with the previous
measurements of $\Gamma_W$ which were done at the CERN $S p \bar{p} S$
collider\cite{UA1,UA2}.
These measurements, which were done at $\sqrt{s}=630$~GeV, obtained:
\begin{eqnarray*}
R & = & 9.5^{+1.1}_{-1.0} \hspace{.54 in}   \Gamma_W = 2.18^{+0.26}_{-0.24}\pm
0.04 \rm  GeV \;\;\;\;        (UA1) \\ R & = & 10.4^{+0.7}_{-0.6}\pm 0.3
\;\;\;  \Gamma_W = 2.10^{+0.14}_{-0.13}\pm 0.09 \rm GeV \;\;\;\;    (UA2) \\
\end{eqnarray*}
The $R$ values from these experiments can be combined to give
$R  = 10.11 \pm 0.59$.
Using a theoretical total cross section ratio\cite{UA2}
of $\sigma_W/\sigma_Z = 3.26 \pm 0.09$ for this center of mass energy
 and using the same values for
$B(\Zll)$ and $\Gamma(\Wlnu)$ as above, these measurements give a combined
value of $\Gamma_W = 2.16\pm 0.14$~GeV.
Finally, the CERN and Fermilab measurements can be combined to obtain a
world average of $\Gamma_W = 2.062\pm 0.059$~GeV.
In forming this average, in order to allow for a correlation of the
errors on the predicted cross section  ratio $\sigma_W/\sigma_Z$ at the two
center of mass energies through the choice of pdf,
the  errors due to the input quantities are combined
linearly, while the experimental errors (i.e. due to $R$) are combined
in quadrature.

The standard model prediction of the $W$ total width is
a function of the $W$ mass, and using the value of $M_W$ given
above we obtain:
\begin{equation}
\Gamma_W = \left( 3 + 6(1 + \alpha_s(M_W)/\pi \right) \cdot \Gamma(\Wlnu)
= 2.077 \pm 0.014 \;\; \rm GeV
\end{equation}
with which the experimental value is in very good agreement.
In the past this comparison of measured and theoretical values of $\Gamma_W$
was used to set a model
independent limit on the mass of the top quark  for the case of $m_t < M_W$.
Given that the top quark is in fact much heavier than the $W$ boson,
this comparison can be used to set an upper limit on the ``excess width'',
$\Delta \Gamma_W \equiv \Gamma_W{\rm (meas)} - \Gamma_W {\rm  (SM)}$,
allowed by experiment for  non--standard model decay processes, such
as decays into supersymmetric charginos and neutralinos\cite{SUSY}, or into
heavy quarks\cite{ALVAREZ}.
Comparing the above world average value of $\Gamma_W$ with the standard
model prediction gives a 95\% CL upper limit of
$\Delta\Gamma < 109$~MeV on unexpected
decays.

While the indirect method gives the most precise measurements
of $\Gamma_W$, there is much interest in performing a direct measurement
from the $W$ lineshape itself.
CDF has performed such an analysis\cite{CDF_DIRECT}  on their 1992--93
electron channel data.
Using the $W$ Breit--Wigner, a fit is performed to the transverse mass
distribution far above
the $W$ pole ($M_T > 110$~GeV)
where the Breit--Wigner tail dominates over the Gaussian resolution of the
detector. A binned log--likelihood fit is performed to this
region of the  $M_T$  distribution and
the data is compared to Monte Carlo generated templates
generated with $0.667 \leq
\Gamma_W \leq 3.667$~GeV in steps of 200~MeV.
Restricted to the tail of the distribution, the
measurement is at present limited by
statistics.
They obtained $\Gamma_W = 2.11 \pm 0.28 \pm 0.16$~GeV,
where the systematic
error (8\%) is dominated by uncertainty in modelling
the $W$ transverse
momentum distribution (6\%)  and the missing \Et\
resolution (5\%).

\subsection*{Summary and Prospects}
{}From their analysis of their 1992--93 data, both \D0 and CDF have obtained
new
measurements of the inclusive vector boson production cross sections.
The ratio of these cross sections is measured to a precision of
$\sim 4-5\%$ by each experiment.
Combining these results with theoretical calculations we obtain a
measurement of $\Gamma_W$ that has an uncertainty of $\sim 60$~MeV ($\sim
3\%$).  This is presently the most precise measurement of $\Gamma_W$. An
independent
method of directly fitting the lineshape is at present limited by
statistics to a $\sim 15$\% uncertainty. The 1994--95 run of the Tevatron
is expected to  yield integrated luminosities of $\geq 100$~\ipb, which will
permit the experiments to
reduce their errors on $\Gamma_W$ to the  $\sim 2\%$ level,
limited by the uncertainties on the  acceptance and efficiency.
Also,
the theoretical cross section ratio, which is limited by pdf uncertainties,
will also make it difficult to significantly improve on this.
By comparison, the direct method has been estimated\cite{CDF_DIRECT}
to result in a $\sim 5\%$  measurement, given a 200~\ipb\ data sample
and combining  results from both experiments.

I am grateful to the \D0 and CDF collaborations for discussion of their data.
We thank the Fermilab Accelerator, Computing, and Research Divisions, and
the support staffs at the collaborating institutions for their contributions
to the success of this work.
This work was supported by
the Director, Office of Energy Research,
Office of High Energy and Nuclear Physics,
Division of High Energy Physics of the U.S.~Department of Energy
under Contract DE-AC03-76SF00098.
\end{document}